\documentclass{aa}  
\usepackage{graphicx}
\usepackage{lscape}
\usepackage{longtable}
\usepackage{natbib}
\usepackage{color}
\usepackage{array} 
\usepackage{tikz,array}
\usetikzlibrary{calc}
\usepackage{multirow}
\usepackage{here}
\usepackage{txfonts}
\usepackage{amsmath,amstext}
\usepackage{epstopdf}
\usepackage{float}
\usepackage{mathtools}
\usepackage{booktabs}
\usepackage{subfigure}
\usepackage{url}
\usepackage{helvet}
\usepackage{tabularx}
\usepackage{multirow}
\usepackage{natbib}
\usepackage[flushleft]{threeparttable}
\usepackage{lscape}
\usepackage{pdflscape}
\usepackage{longtable}
\usepackage{wasysym}
\usepackage{float}
\usepackage{relsize}
\usepackage{color}
\usepackage{breqn}
\usepackage{bm}
\usepackage{enumitem}
\usepackage{makecell}

\usepackage[colorlinks, citecolor=blue, linkcolor=blue]{hyperref}
\hypersetup{colorlinks,breaklinks, linkcolor=blue,urlcolor=magenta, anchorcolor=blue,citecolor=blue}
\bibpunct{(}{)}{;}{a}{}{,} 

\newcommand{\be}{\begin{equation}}
\newcommand{\ee}{\end{equation}}

\def\gcm3{\hbox{g cm$^{-3}$}}       

\newcommand{\astrasens}{\texttt{astrasens}}
\newcommand{\gaia}{\textit{Gaia}}
\newcommand{\kepler}{\textit{Kepler}}

\usepackage{graphicx}
\usepackage{txfonts}
\begin{document}

   \title{The AstraLux-TESS high spatial resolution imaging survey\thanks{Based on observations collected at the Centro Astron\'omico Hispano en Andaluc\'ia (CAHA) at Calar Alto (on proposal IDs H18-2.2-011 and F19-2.2-007 with PI: D. Barrado; as well as proposal IDs H19-2.2-005, F20-2.2-014, H20-2.2-005,  21B-2.2-009, 22A-2.2-009, 22B-2.2-004, 24A-2.2-004 with PI: J. Lillo-Box), operated jointly by Junta de Andaluc\'ia and Consejo Superior de Investigaciones Cient\'ificas (IAA-CSIC)}}

   \subtitle{Search for stellar companions of 215 planet candidates from TESS}

   \author{
J.~Lillo-Box\inst{\ref{cab}}, 
M.~Morales-Calder\'on\inst{\ref{cab}},
D.~Barrado\inst{\ref{cab}},
O.~Balsalobre-Ruza\inst{\ref{cab}},
A.~Castro-Gonz\'alez\inst{\ref{cab}},\\
I.~Mendigut\'ia\inst{\ref{cab}},
N.~Hu\'elamo\inst{\ref{cab}},
B.~Montesinos\inst{\ref{cab}},
M.~Vioque\inst{\ref{eso}}
}

\titlerunning{The Astralux-TESS survey}
\authorrunning{Lillo-Box et al.}

\institute{
Centro de Astrobiolog\'ia (CAB), CSIC-INTA, ESAC campus, Camino Bajo del Castillo s/n, 28692, Villanueva de la Ca\~nada (Madrid), Spain \email{Jorge.Lillo@cab.inta-csic.es}
\label{cab} 
\and 
European Southern Observatory, Karl-Schwarzschild-Strasse 2, D-85748 Garching bei M\"unchen, Germany \label{eso} 
}

   \date{Accepted April-2024}

 
  \abstract
   {Chance-aligned sources or blended companions can cause false positives in planetary transit detections or simply bias the determination of the candidate properties. In the era of high-precision space-based photometers, the need for high spatial resolution images has been demonstrated to be critical for validating and confirming transit signals. This already applied to the \textit{Kepler} mission, is now applicable to the TESS survey, and will be critical for the PLATO mission. }
   {In this paper we present the results of the AstraLux-TESS survey, a catalog of high spatial resolution images obtained with the AstraLux instrument at the Calar Alto observatory (Almer\'ia, Spain) in the context of the TESS Follow-up Observing Program.}
   {We used the lucky imaging technique to obtain high spatial resolution images from planet candidate hosts included mostly in two relevant regimes: exoplanet candidates belonging to the level one requirement of the TESS mission (planets with radii $R<4~R_{\oplus}$) and TESS planet candidates around intermediate-mass main-sequence stars.}
   {Among the 185 planet host candidate stars observed, we found 13 (7\%) to be accompanied by additional sources within a separation of {2.2 arcsec}. Among them, six are not associated with sources in the \gaia\ DR3 catalog, thus contaminating the TESS light curve. Even if no contaminants have been detected, we can provide upper limits and probabilities to the possible existence of field contaminants through the sensitivity limits of our images. Among the isolated hosts, we can discard hazardous companions (bright enough to mimic a planetary transit signals) with an accuracy below 1\% for all their planets.}
   {The results from this catalog are key to the statistical validation of small planets (prime targets of the TESS mission) and planets around intermediate-mass stars in the main sequence. These two populations of planets are difficult to confirm with the radial velocity technique because of the shallow amplitude of small planets and the high rotational velocities and low number of available spectral lines in the intermediate stellar mass regime. Our results also demonstrate the importance of this type of follow-up observation for future transit missions such as PLATO, even in the \gaia\ era.}

   \keywords{Planets and satellites: general -- Techniques: high angular resolution -- Surveys   }

   \maketitle
%
\section{Introduction}
Confirming an exoplanet candidate detected through observations from a single technique is a chimera. The main detection methodologies rely on several indirect effects of the presence of planets around other stars. The two most efficient methods for planet detection to date (transits and radial velocity) suffer from strong degeneracies and unknowns that by themselves can only provide candidates rather than bona fide confirmed planets. While the radial velocity method can only provide a minimum mass for the orbiting object, the transit technique (especially from large space-based missions) cannot by itself distinguish a planet transit from other non-planetary configurations (e.g., \citealt{daemgen09, morton12, lillo-box12}). Additionally, although astrometry is capable of fully characterizing the system parameters and thus confirming the planetary nature, its inability to reach the inner regions of planetary systems still makes it an inefficient technique for completing the exoplanet population. 

Since the launch of the \kepler\ mission in 2009 (\citealt{borucki03}) and other high-precision photometers put into space, the need for ground-based follow-up programs has become clear. Among other types of observations, the need for high spatial resolution images has been demonstrated to be crucial to confirm the nature of the transiting candidates (and other types of signals such as ellipsoidal variations; see, e.g., \citealt{lillo-box14}). In the era of the Transiting Exoplanet Survey Satellite (TESS; \citealt{ricker14}) and in view of the upcoming PLAnetary Transits and Oscillation of stars (PLATO) mission (\citealt{rauer14}), where the pixel size of the cameras is 21$\times$21 arcsec for TESS and 15$\times$15 arcsec for PLATO, this type of follow-up image of candidate planet hosts will continue to be crucial. Fortunately, the launch of \gaia\ (\citealt{gaia}) and its different data releases (\citealt{gaia18,gaiaDR3}) have greatly helped in the identification of stellar companions with good completeness down to {separations of more than 0.7~arcsec (e.g., \citealt{ziegler18,michel24})}. This, together with such contamination identification tools as \texttt{tpfplotter}\footnote{Publicly available at \url{https://github.com/jlillo/tpfplotter}.} \citep{aller20}, has served to focus the high spatial resolution programs to the regions very close to the planet host candidates (i.e., below 2 arcsec).

The utility of high spatial resolution images is then twofold. On the one hand, the identification of nearby companions has critical implications in the planetary candidacy of a transiting signal (sometimes even challenging its planetary nature; see, e.g., \citealt{lillo-box21b}). For instance, eclipsing binaries blended by a brighter star {can} produce a transit depth compatible with that of a planet-sized object. But even in the event that the planet is actually real and orbits the main bright target, the additional source contaminates and biases the determination of the physical properties of the planet if not accounted for (see, e.g., \citealt{daemgen09}). Indeed, in the TESS mission, the light curves processed by the Science Processing Operations Center (SPOC; \citealt{jenkins16}) include among the different corrections a crowding correction to account for contaminant sources appearing in the \gaia\ catalogs in order to produce the Pre-search Data Conditioning SAP flux (PDCSAP). On the other hand, when no contaminant sources are found, the estimation of the sensitivity limits of the image are key to inferring the probability that no contaminant sources are present and, even further, discarding a given probability that no other sources capable of mimicking the planetary transit are present (see \citealt{lillo-box14b}). This second outcome of high spatial resolution images is a key step in the so-called exoplanet statistical validation process (e.g., \citealt{morton11,torres15,santerne15,diaz14b}), where a probability is given that the transit signal does not come from other possible non-planetary configurations. 

In this regard, there are specific configurations that can highly benefit from this type of observation. First of all, small planet candidates with shallow transits could actually be the signal of faint eclipsing binaries blended with the potential bright host star (the so-called blended eclipsing binary scenario). This is because the shallowness of the transit implies that even faint (faraway or low-mass) eclipsing binaries blended with bright sources could mimic such transit depths. In these cases, high spatial resolution images are critical to validating the signal as being planetary. Indeed, this has been the case in many studies both from \kepler\ (e.g., \citealt{barclay13,armstrong21}), K2 (e.g., \citealt{castro-gonzalez22}), and TESS (e.g., \citealt{mantovan22,giacalone22}). In the case of TESS, this is even more relevant because its primary goal is the confirmation of the so-called Level One Science Requirement (LOSR) planets, that is, confirming at least 50 planets smaller than 4~R$_{\oplus}$. Second of all, despite theories of planet formation foreseeing the genesis of a large population of planets around intermediate-mass stars, given the huge amount of available material in their forming disks \citep[e.g.,][]{mulders15,stapper23,guzman-diaz23}, very few have actually been detected and truly confirmed (that is, with measured planet masses), especially in close orbits (P$_{\rm orb}<100$~days). Indeed, in total, among the more than 5500 planets confirmed or validated so far, only 42 of them revolve around main-sequence stars hotter than spectral type F (i.e., T$_{\rm eff}>7200$ K). Furthermore, among these, only around 20 revolve at periods below 100 days. The reason is clear: the great difficulty of confirming transiting candidates with the radial velocity technique given the lack of spectral line content and the broadened spectral features due to the high rotational velocities of these stars. Consequently, high spatial resolution images are, again, key to facilitating the validation of this population of planets that are fundamental to understanding planet formation theories in different stellar regimes. 

In this paper, we present a catalog of 185 TESS Objects of Interest (TOIs) hosting 215 planet-like transiting signals that we have been observing in high spatial resolution since the beginning of the TESS mission using the AstraLux instrument at the Calar Alto Observatory (Almer\'ia, Spain). In Sect.~\ref{sec:Observations}, we describe the survey properties and the target sample. In Sect.~\ref{sec:analysis}, we use those observations to detect additional nearby companions and quantify the likelihood of the planet host stars to have blended companions with capabilities of mimicking the transit signal. Finally, in Sect.~\ref{sec:conclusions}, we provide the conclusions of the observations and the foreseen need for more high spatial resolution imaging surveys for subsequent high-precision photometry missions, including PLATO.

\section{The AstraLux-TESS survey}
\label{sec:Observations}

In a manner similar to our follow-up of the \kepler\ planet candidates (see \citealt{lillo-box12,lillo-box14b}), since the launch of the TESS mission and the delivery of the first exoplanet candidates in 2018, we have developed a high spatial resolution observing program of these candidate planet hosts by using the AstraLux instrument installed at the 2.2m telescope of the Centro Astron\'omico Hispano en Andaluc\'ia (CAHA), Calar Alto observatory (Almer\'ia, Spain). In the period 2018-2022, around five to ten nights per semester were devoted to this program.

\begin{figure*}[h]
\centering
\includegraphics[width=1\textwidth{}]{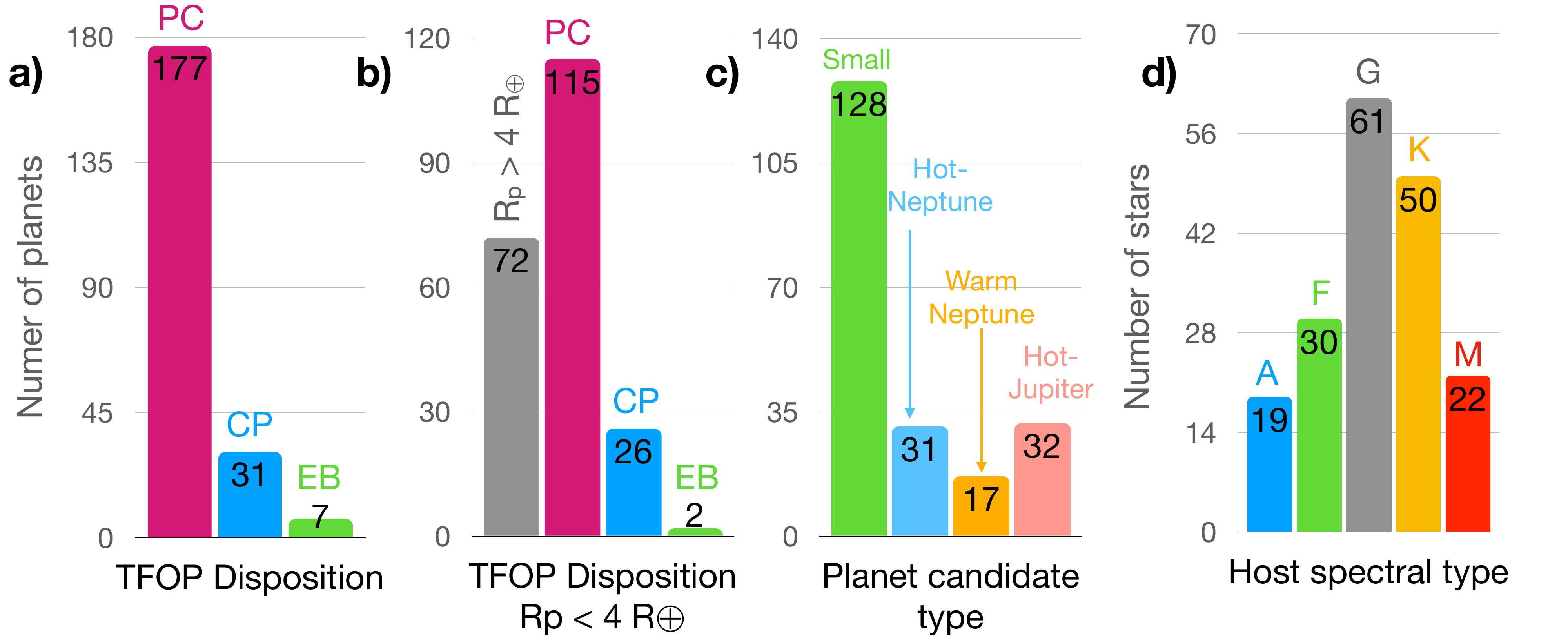}
\caption{Summary of the TESS Follow-up Program disposition (status) of the planet candidates around the observed targets in the AstraLux-TESS survey presented in this work. \textbf{a)} Disposition of all planets in the sample. \textbf{b)} Disposition of the planets with $R_p<4R_{\oplus}$. \textbf{c)} {Distribution of the planet type defined as small planets below the Neptune desert, hot-Neptunes in the desert, Neptune-like planets beyond the desert (warm-Neptunes), and hot-Jupiters. \textbf{d)} Distribution of the host stars per spectral type up to T$_{\rm eff}=10\,000$~K. Three targets are above this threshold and are thus not included in the diagram).}} 
\label{fig:sampleprop}
\end{figure*}

\begin{figure*}
\centering
\includegraphics[width=0.48\textwidth{}]{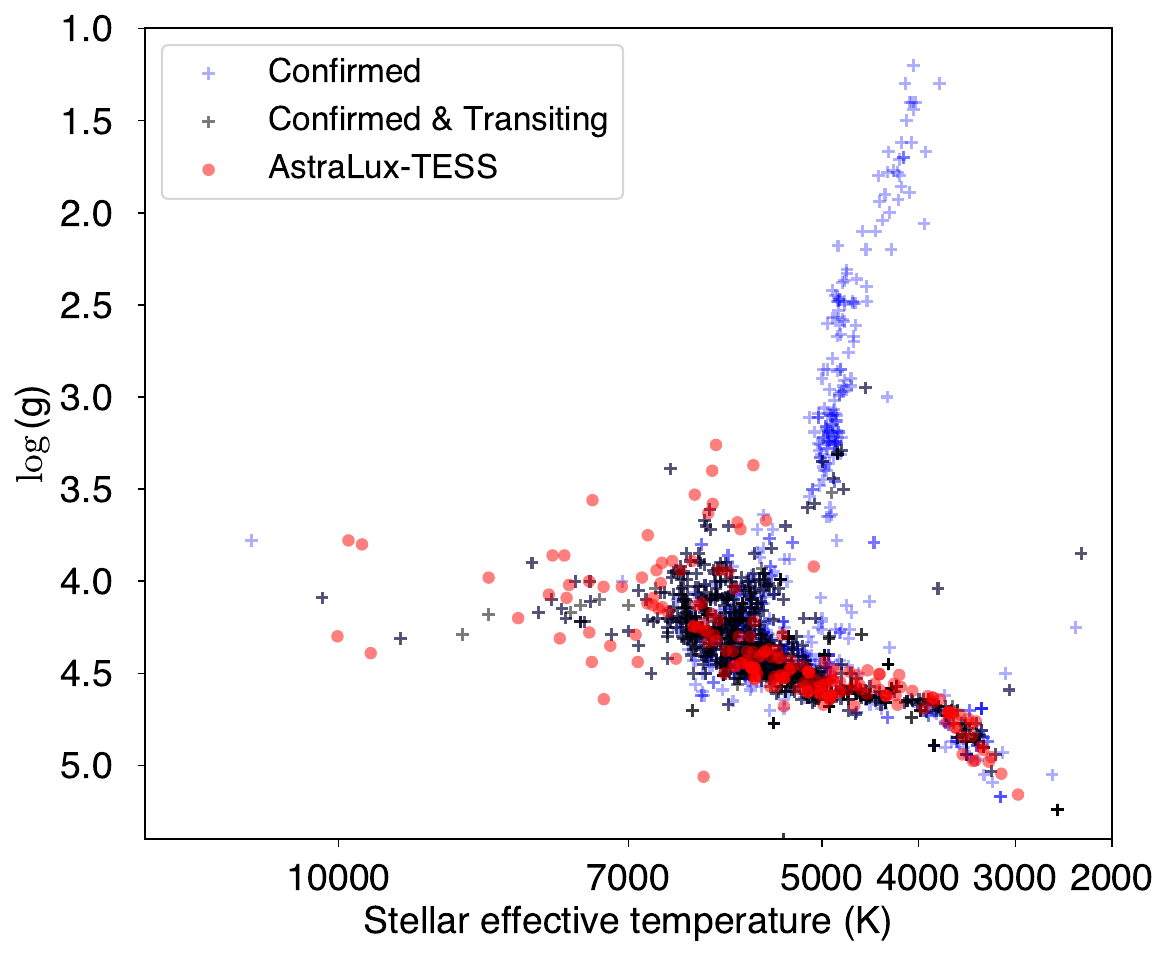}
\includegraphics[width=0.48\textwidth{}]{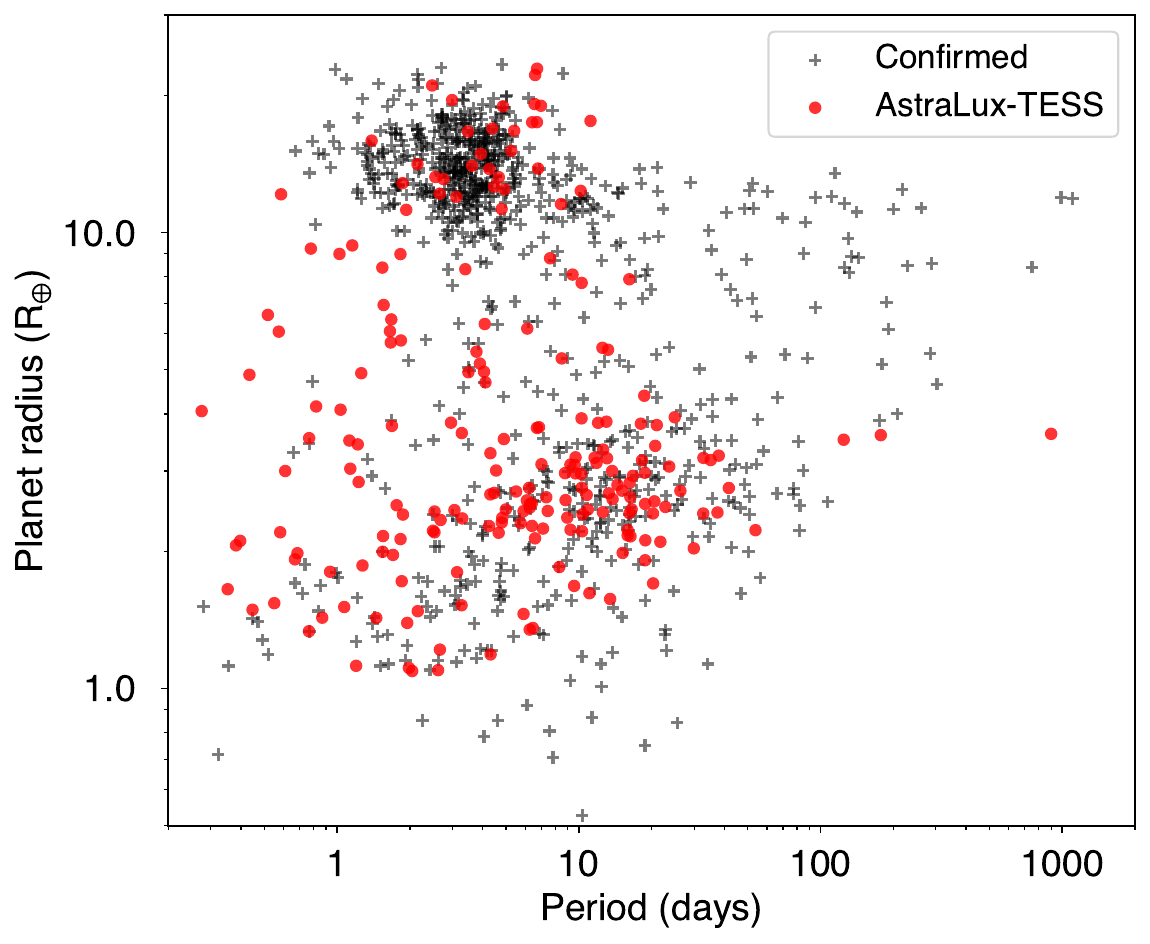}
\caption{Properties of the sample of TOI host stars {and planet candidates} observed within the AstraLux-TESS survey of high spatial resolution images. \textbf{Left:} Stellar surface gravity ($\log{g}$) against effective temperature of the candidate host stars (according to the TESS Input Catalog v8.2, \citealt{TIC82}) with planets with measured masses (blue crosses), confirmed transiting planets (black crosses), and the sample of candidate hosts observed within the AstraLux-TESS survey. \textbf{Right:} Planet radius versus orbital period of all transiting-confirmed planets (black crosses) and the sample of candidate hosts observed with AstraLux (red circles).}
\label{fig:sample}
\end{figure*}

The target selection for each campaign was mainly driven by the LOSR of the TESS mission, envisioned to confirm at least 50 planets smaller than 4~R$_{\oplus}$. With high priority, we pointed toward sources listed as planet candidate hosts in the TOI data release.\footnote{The TOI releases are published on the following website: \url{https://tev.mit.edu/data/}.} In total, 185 stars hosting 215 planet candidates were observed in this survey (see Table~\ref{tab:sample}). In Fig.~\ref{fig:sampleprop}, we show the main properties of the sample of planet candidates and host stars observed in this AstraLux-TESS survey. Among the sample, according to the current classification from the TESS Follow-up Observing Program (TFOP), {as of January 2024}, 31 have a confirmed planet (CP) disposition, seven are classified as eclipsing binaries (EB), and the 177 left remain as planet candidates (PCs; see panel $a$ in Fig.~\ref{fig:sampleprop}). In total, 143 among the 215 TOIs within our observed targets have an estimated planetary radius below 4~$R_{\oplus}$, fulfilling the criterion to be LOSRs of the TESS mission. However, 26 of them have already been confirmed or validated as planets, and two are classified as eclipsing binaries, thus leaving 115 that still have a planet candidate disposition (see panel $b$ of Fig.~\ref{fig:sampleprop}). Concerning the planet candidate properties (see right panel from Fig.~\ref{fig:sample}), we observed the host stars of 32 hot-Jupiter candidates, 17 warm-Neptune candidates, 31 hot-Neptune candidates in the desert, and 128 small planet candidates (see panel $c$ of Fig.~\ref{fig:sampleprop}). Added to this, we also included planets from different stellar regimes (see left panel from Fig.~\ref{fig:sample}), especially planets around slightly evolved stars potentially in the subgiant phase\footnote{{This is part of the Temporal Evolution and Metamorphosis of exoPlanets and their atmOspheres (TEMPO) project. See \url{https://remote-worlds-lab.cab.inta-csic.es/project_TEMPO.html}.}} (10 TOIs) as well as one planet candidate around a white dwarf and planets around massive A-type main-sequence stars (with $T_{\rm eff} >7200$~K; 19  stars in total) focused on the intermediate-mass regime (see Mendigut\'ia et al., 2024, submitted). The remaining targets are main-sequence stars of FGKM spectral types (see panel $d$ of Fig.~\ref{fig:sampleprop}). Figure~\ref{fig:sample} shows the properties of the observed candidate host stars (left panel) and of their transiting planet candidates (right panel). Some of our targets belong to intensive radial velocity follow-up programs with CARMENES (CARM-TESS, PI: E. Pall\'e), CAFE (CAFE-TESS, PI: J. Lillo-Box), and HARPS (NOMADS, PI: D. Armstrong), {providing a full set of follow-up observations that typically allow for a definitive confirmation of the exoplanetary nature (see \citealt{lillo-box15PhD})}. 

Among the sample presented in this paper, 31 planets have already been confirmed or validated in other works (e.g., \citealt{hedges21,eisner21}), most of which actually make use of the high spatial resolution images presented here (e.g., \citealt{demory20,turtelboom22}). However, 177 still lack a definitive confirmation or validation analysis. The catalog and sensitivity curves presented here can be used by the community for the latter purpose. Table~\ref{tab:sample} summarizes the main properties of the observed host stars, including their TOI and TESS Input Catalog (TIC) identifications, coordinates, and TESS magnitude (Tmag), and the disposition, orbital period, and radius of their planet candidates.

The survey was designed to reach an average magnitude contrast ($\Delta m$) in the Sloan Digital Sky Survey (SDSS) filter z (hereafter SDSSz) of 5 mag at 0.2 arcsec. However, we used the transit depth of the shallowest planet candidate found around each of the targets to adapt the exposure time in order to reach the maximum magnitude contrast that a given blended source should have to be able to mimic the depth of such a planetary transit. This is given, as explained in previous works such as \cite{lillo-box14b}, by the equation

\begin{equation}
\label{eq:contrast}
        \Delta m_{\rm max} = m_{\rm EB} - m_{\rm target} = -2.5\log{(\delta)},
\end{equation}

\noindent where $\delta$ is the transit depth of the smallest planet candidate in the system. However, ensuring this contrast level is not always possible given the shallowness of the transit depth of some planets. For instance, a 4~R$_{\oplus}$ planet transit around a solar-like star would produce a transit depth of $\delta=1.3$~ppt, implying a contrast of 7 magnitudes. The lucky imaging technique consists of obtaining thousands of short exposure frames to select a small percentage obtained under the best atmospheric conditions (the so-called Strehl ratio; \citealt{strehl1902}). To overcome the variations in the atmospheric layers, one needs to integrate each frame with times below the coherent time (which is typically around 50-100 ms). In most of the cases, for TESS targets, we can integrate with the shortest exposure time possible in AstraLux, which is 10 ms. However, to reach a 7-magnitude contrast in a target with $m_{\rm SDSSz}=10$~mag, we needed to obtain more than 65\,000 frames of 10 ms to reach $m_{\rm SDSSz}=17$~mag for a 10\% frame selection rate\footnote{{This is the optimal selection rate trade-off according to the pipeline developer Felix Hormuth (private communication), and it is recommended in the user's manual of the instrument. See also \cite{hormuth08}.}} (i.e., the 10\% best images are stacked). For a full frame capability of AstraLux (corresponding to a field of view of 24x24 arcsec), this implies a heavy 4 GB file to produce a single image. The situation is obviously worse for shallower transits and fainter targets. Hence, this technique generates a huge amount of heavy files whose storage and processing require very specific capabilities. As a consequence, for very bright sources, we had to truncate the desired maximum contrast magnitude due to technical limitations. 

Overall, for each target we integrated with a given individual exposure per frame, T$_f$, for a total number of frames N$_f$. {These values were chosen based on the target brightness, the requirement of reaching a contrast of 5 mag to discard most of the possible false positives, and prioritizing the shortest possible individual exposure time of 10 ms to reach the best angular resolution.}. Table~\ref{tab:observations} shows the setup for each of the observed images, including the {date}, airmass, number of frames, exposure time of the individual frames, and effective exposure time of the image (that is, for the 10\% selection rate, $T_{\rm eff} = 0.1 \times N_f \times T_f$). We used the SDSSz filter, as it produces the best results with the instrument; it was tested by our team in previous works, but it is also recommended in the instrument user manual (\citealt{hormuth08}). For practical reasons, given that sources separated by more than 2 arcsec can already be detected by the \gaia\ mission, we trimmed the field of view to a $6\times6$ arcsec window. This allowed us to save disk space and computation time while also providing a field of view of 3 arcsec around the target when properly centered. 

Each data cube was then processed by the automatic pipeline at the observatory, which performs the basic data reduction (bias subtraction, flat fielding, and background removal) and subsequently measures the Strehl ratio of each individual image. Then, the best frames according to this parameter were selected, aligned, and stacked to build a final high spatial resolution image of the target. This process is described in detail in \cite{hormuth08} and \cite{lillo-box12}. The reduced images have been uploaded to the Exoplanet Follow-up Observing Program (ExoFOP) repository.\footnote{\url{https://exofop.ipac.caltech.edu/tess/}}

 \begin{table*}
 \setlength{\extrarowheight}{3pt}
 \centering
\caption{General properties of the sample of planet candidates whose host star has been observed with AstraLux. A complete version of this table (including the 185 stellar host candidates) is included in the electronic version of this paper. }
\label{tab:sample}
\begin{tabular}{rrrcrrrrr}
\hline \hline
  \multicolumn{1}{r}{TOI} &
  \multicolumn{1}{r}{TIC} &
  \multicolumn{1}{r}{Candidate} &
  \multicolumn{1}{c}{Candidate} &
  \multicolumn{1}{r}{RA} &
  \multicolumn{1}{r}{DEC} &
  \multicolumn{1}{r}{Tmag} &
  \multicolumn{1}{r}{$P_{\rm orb}$} &
  \multicolumn{1}{r}{$R_{\rm p}$}  \\
  
  \multicolumn{1}{r}{} &
  \multicolumn{1}{r}{} &
  \multicolumn{1}{r}{ID} &
  \multicolumn{1}{c}{disposition} &
  \multicolumn{1}{r}{(deg)} &
  \multicolumn{1}{r}{(deg)} &
  \multicolumn{1}{r}{(mag)} &
  \multicolumn{1}{r}{(days)} &
  \multicolumn{1}{r}{($R_{\oplus}$)} \\

\hline
  238 & 9006668 & 238.01 & PC & 349.231362 & -18.606628 & 9.897 & 1.273099 & 1.86 \\
  442 & 70899085 & 442.01 & CP & 64.1900021 & -12.084015 & 10.733 & 4.0520368 & 4.95 \\
  460 & 9804616 & 460.01 & EB & 47.6431 & -9.2768 & 14.121 & 0.516834 & 6.59 \\
  462 & 420049884 & 462.01 & PC & 36.276352 & 1.323725 & 10.605 & 4.1075011 & 4.69 \\
  464 & 398733009 & 464.01 & PC & 32.16976 & -9.046048 & 15.722 & 0.818567 & 4.15 \\
  488 & 452866790 & 488.01 & CP & 120.5953 & 3.3388 & 11.13 & 1.197995 & 1.12 \\
  494 & 19519368 & 494.01 & PC & 124.69586 & -6.785861 & 10.039 & 1.7018818 & 1.96 \\
  498 & 121338379 & 498.01 & PC & 129.09088 & -3.860234 & 9.527 & 0.275047 & 4.05 \\
  515 & 366622912 & 515.01 & PC & 128.067122 & 11.630687 & 13.725 & 3.114891 & 11.97 \\
  521 & 27649847 & 521.01 & PC & 123.3438 & 12.2217 & 12.429 & 1.5428518 & 1.99 \\
  ... & ...&... &... &... &... &... &... &  \\
\hline
\end{tabular}
\end{table*}

\begin{table*}
\setlength{\extrarowheight}{3pt}
\centering
\caption{Observing conditions and setup for the observed targets. A complete version of this table is included in the electronic version of this paper. }
\label{tab:observations}
\begin{tabular}{ccccccc}
\hline \hline
TOI & Night & MJD\tablefootmark{$^{\dagger}$} & Airmass & N$_f$ & T$_{f}$ & T$_{\rm eff}$ \\
 & YYYY-MM-DD & (days) & & (frames) & (ms) & (s) \\  \hline
238 & 2019-08-13 & 8709.5164 & 2.01 & 39472 & 20 & 78.9 \\
442 & 2020-02-25 & 8905.3439 & 1.87 & 1267 & 20 & 2.5 \\
460 & 2019-08-13 & 8709.6333 & 1.98 & 8109 & 60 & 48.7 \\
462 & 2019-08-13 & 8709.5883 & 1.73 & 40000 & 20 & 80.0 \\
464 & 2019-08-13 & 8709.6759 & 1.47 & 955 & 60 & 5.7 \\
488 & 2019-05-07 & 8611.3658 & 1.98 & 7711 & 10 & 7.7 \\
494 & 2019-05-06 & 8610.3817 & 2.58 & 10000 & 10 & 10.0 \\
498 & 2019-05-07 & 8611.331 & 1.59 & 10000 & 10 & 10.0 \\
515 & 2019-05-07 & 8611.3352 & 1.32 & 1000 & 100 & 10.0 \\
521 & 2019-05-07 & 8611.3826 & 1.81 & 18000 & 50 & 90.0 \\
522 & 2019-05-06 & 8610.3993 & 3.29 & 987 & 10 & 1.0 \\
544 & 2021-03-23 & 9297.3841 & 2.19 & 6936 & 10 & 6.9 \\
  ... & ...&... &... &... &... &...  \\
\hline
\end{tabular}
\tablefoot{\\
\tablefoottext{$^{\dagger}$}{Modified Julian date, defined as the Julian date minus 2450000.}\\
}
\end{table*}

\section{Analysis and results}
\label{sec:analysis}

We analyzed the high spatial resolution images in three steps. First, we looked for close companions in the image. Second, we obtained the sensitivity curves, which provided an estimation of the maximum contrast magnitude of the image at different separations that can be detected. Third, we used the sensitivity curve to obtain the probability of the presence of a blended source (and, more specifically, a blended eclipsing binary capable of mimicking the transit depth) not detected in our image. The first two steps were done on our own built, publicly available code \texttt{astrasens},\footnote{\url{https://github.com/jlillo/astrasens}} and the third step was performed using our publicly available code \texttt{bsc}.\footnote{\url{https://github.com/jlillo/bsc}} These three steps and their results are described in the following sections.

\subsection{Source identification}
We first identified the TOI in the AstraLux-extracted image obtained as described in the previous section. This was done through a preliminary peak identification and source analysis using the \texttt{photutils} package \citep{photutils}. We chose the brightest source in the image to be the target source corresponding to the planet candidate host. We note that all companions detected in this sample are significantly fainter than the main target. Once the TOI was identified, we fit, through a maximum likelihood approach, its point spread function (PSF) by using a combination of a two-dimensional Lorentzian and a Gaussian, both centered in the same pixel and with variable widths (free parameters of the modeling process). This PSF clearly provides the best description of the AstraLux PSF (as also stated in \citealt{hormuth08}). We then created a model image based on this PSF and subtracted it from the original image. The residual image was then used for a second source identification in a search for additional companions. The sources identified in this second step were then triaged based on their roundness property and minimum flux (we required a detection above 5$\sigma$, where $\sigma$ here defines the median sky background of the residual image). 

For targets with companions, we measured the magnitude contrast by using the \texttt{aperture_photometry} module of the \texttt{photutils} package. We assumed a circular aperture of 5 pixels and an annulus at an 8-pixel separation with a width of 3 pixels. We then measured the median flux per pixel in the annulus and subtracted it from the per pixel flux inside the photometric aperture. After the background correction, the sum of the flux pixels was then used to estimate the uncalibrated magnitude for each source. The contrast magnitude of the companions was then determined by subtracting the uncalibrated magnitude of the target. Table~\ref{tab:companions} shows the detected companions and their properties. {The separation, position angle from the north direction, and relative contrast magnitude against the target of the identified sources were then stored by \astrasens\ in an output file.}

\begin{figure*}
\centering
\includegraphics[width=0.85\textwidth{}]{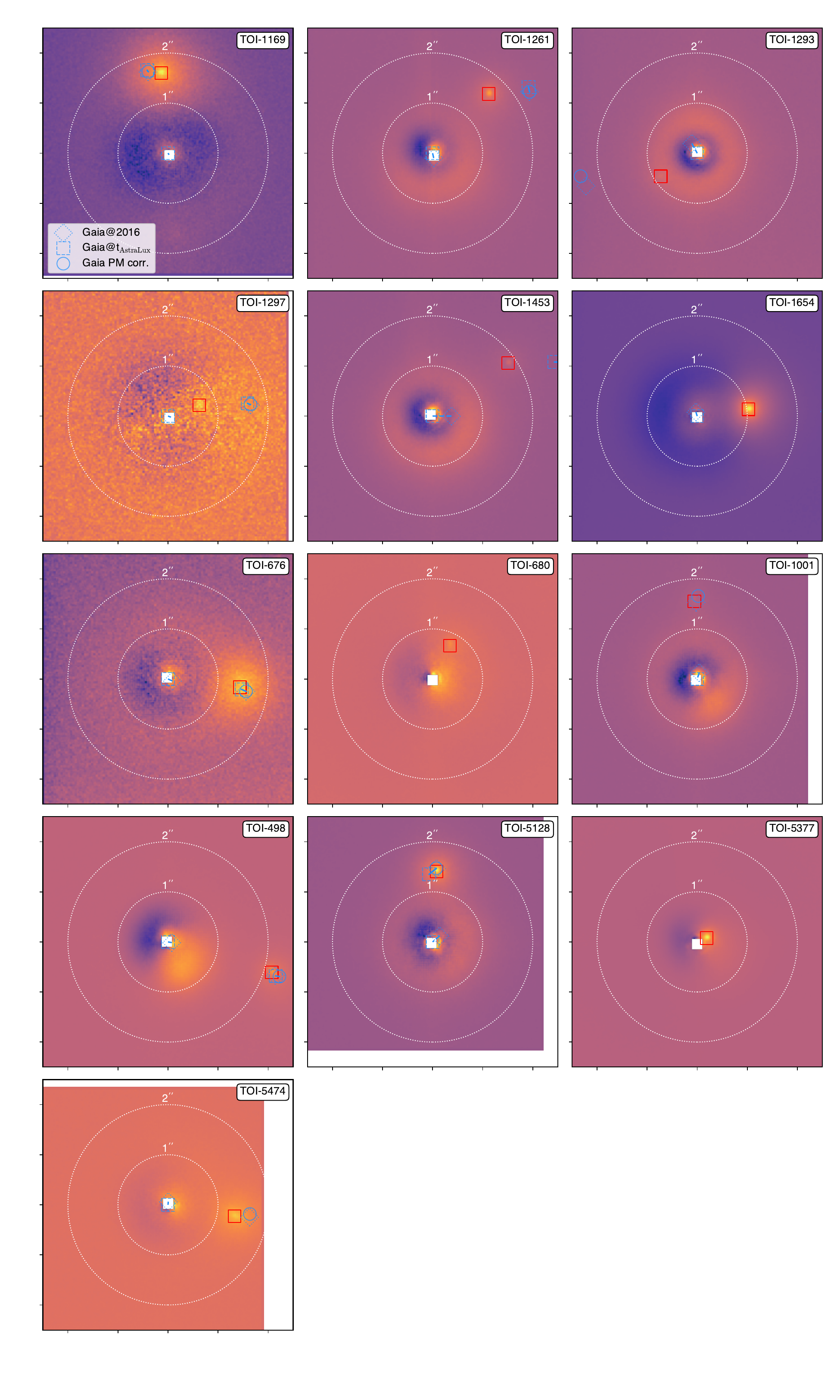}
\caption{AstraLux images of the TOIs from our sample with detected companions closer than {2.2 arcsec}. The TOI is located in the center of the image, and the detected companions are marked with open red squares. The \gaia\ DR3 sources in the field are marked with blue symbols: dotted diamond for the 2016 position, dashed square for the position corrected from its proper motion at the time of the AstraLux observation, and solid circle for the position after also correcting for the proper motion of the TOI. North is up and east is left in all images.}
\label{fig:companions}
\end{figure*}

In total, among the 185 targets observed, we have identified {20 companions around 20 different stars}. {The properties of these companions are shown in Table~\ref{tab:companions}}. Among them, 13 (7\%) are within {2.2 arcsec} of the primary TOI target (see Fig.~\ref{fig:companions}). Figure~\ref{fig:companions} shows the AstraLux images of the 13 targets with companions detected closer than {2.2 arcsec}, indicating the location of the \gaia\ sources in the field. We note that the locations of the \gaia\ sources in the field have been corrected for the proper motion of the source and the proper motion of the main target at the time of the AstraLux observation. Interestingly, we found six companions that perfectly match the expected location of the \gaia\-identified sources (with a 0.3 arcsec margin). We then associated the detected companions to the corresponding \gaia\ identifications. These sources are marked with a dagger symbol ($^{\dagger}$) in Table~\ref{tab:companions}. Additionally, in four cases, the \gaia\ catalog reports the presence of companions that do not match the location of the source detected in our AstraLux images, and more importantly, our image does not show any emission at the expected \gaia\ location after correcting for proper motions. These companions are marked with a double dagger symbol ($^{\dagger\dagger}$) in Table~\ref{tab:companions}.\footnote{We note that this table includes companions beyond 2 arcsec.} The mismatch between the AstraLux location of the source and the expected position from \gaia\ can either be due to an error in the determination of the proper motions or coordinates for such nearby faint sources or an orbital motion, especially in the case of sources at distances similar to TOI-1450 or TOI-1452. Finally, three sources (TOI-680, TOI-2072, and TOI-5377) show no \gaia\ counterpart at all and are thus new companions previously undetected.

\begin{table*}
\setlength{\extrarowheight}{3pt}
\caption{Detected companions in the sample of stars observed with AstraLux presented in this work. The main target is labelled with an "A" in the ID column, AstraLux-detected companions are labelled with a "B," and other \gaia\ sources are labelled "\textit{Gaia}."} 
\label{tab:companions}
\tiny
\begin{tabular}{lccccccccc}
\hline \hline
TOI & ID & sep & PA & $\Delta$m$_{\rm SDSSz}$ & \gaia\ DR3 counterpart  & d & $\mu_{\alpha}$ & $\mu_{\delta}$ \\
    &    & (arcsec) & (deg.) & (mag) &  & (pc)& (mas/yr) & (mas/yr)\\
\hline

TOI-498 & A & - & - & - & 3071584413361857280 & $192.1\pm 1.6$ & -21.69 & 4.02 \\
 & B & $2.1792\pm 0.0029$ & $106.82\pm 0.18$ & $3.355\pm 0.024$ & 3071584417657701504 & $5480 \pm 910$ & 2.74 & -2.57 \\ \hline

TOI-676\tablefootmark{$^{\dagger}$} & A & - & - & - & 3546125626089939968 & $709\pm 11$ & -20.85 & 9.13 \\
 & B & $1.4726\pm 0.0087$ & $98.1\pm 1.5$ & $3.42\pm 0.23$ & 3546125630386919424 & $590\pm 23$ & -20.76 & 8.73 \\ \hline

TOI-680 & A & - & - & - & 3574891599052022144 & $160.1\pm 1.7$ & -56.67 & -6.57 \\
 & B & $0.7679\pm 0.0045$ & $26.59\pm 0.14$ & $3.690\pm 0.042$ & - & - & - & - \\ \hline

TOI-1001 & A & - & - & - & 3067555424799970560 & $287.5\pm 1.5$ & -4.92 & -15.43 \\
 & B & $1.5541\pm 0.0066$ & $-1.43\pm 0.14$ & $5.030\pm 0.075$ & 3067555429095077376 & - & - & - \\ \hline

TOI-1169\tablefootmark{$^{\dagger}$} & A & - & - & - & 1122163457093871744 & $258.50\pm 0.97$ & -5.67 & 7.79 \\
 & B & $1.6507\pm 0.0030$ & $-5.840\pm 0.062$ & $1.285\pm 0.039$ & 1122163461390307072 & - & - & - \\
 & \textit{Gaia} & 1.67 & -13.8 & 0.7 & 1122163461390307072 & $257.3\pm 1.2$ & -3.91 & 7.99 \\  \hline

TOI-1201\tablefootmark{$^{\dagger}$} & A & - & - & - & 5157183324996790272 & $37.636\pm 0.032$ & 164.07 & 46.55 \\
 & B & $8.2994\pm 0.0017$ & $-99.310\pm 0.032$ & $1.505\pm 0.036$ & 5157183324996789760 & $37.680\pm 0.033$ & 174.43 & 45.46 \\ \hline

TOI-1242\tablefootmark{$^{\dagger}$} & A & - & - & - & 1624420987137541376 & $109.52\pm 0.16$ & 1.64 & -39.84 \\
 & B & $4.3836\pm 0.0048$ & $3.776\pm 0.030$ & $2.855\pm 0.050$ & 1624420991433145472 & $109 \pm 22$ & 1.84 & -39.60 \\ \hline

TOI-1261\tablefootmark{$^{\dagger\dagger}$} & A & - & - & - & 1417009251113227648 & $199.90\pm 0.44$ & -4.12 & 18.08 \\
 & B & $1.6346\pm 0.0041$ & $42.088\pm 0.070$ & $2.685\pm 0.037$ & 1417009255408193536\tablefootmark{$^{\ddagger}$} & - & - & - \\ 
 & \textit{Gaia} & 2.29 & 57.52 & 3.2 & 1417009255408193536 & $201.2\pm 1.4$ & -3.32 & 17.07 \\ \hline

TOI-1280 & A & - & - & - & 2240116441786134912 & $92.325\pm 0.080$ & 19.18 & -23.27 \\
 & B & $6.2683\pm 0.0018$ & $88.02\pm 0.19$ & $-1.230\pm 0.029$ & 2240116441786135680 & $672.3\pm9.0$ & -6.35 & -8.84 \\ \hline

TOI-1293\tablefootmark{$^{\dagger\dagger}$} & A & - & - & - & 1640813919530112512 & $236.58\pm 0.91$ & 20.73 & -37.98 \\
 & B & $0.8897\pm 0.0067$ & $236.57\pm 0.29$ & $4.52\pm 0.15$ & 1640813919529592448\tablefootmark{$^{\ddagger}$} & - & - & - \\
 &\textit{Gaia}  & 2.32 & 253.6 & 4.7 & 1640813919529592448 & - & - & - \\ \hline

TOI-1297\tablefootmark{$^{\dagger\dagger}$} & A & - & - & - & 1679498552524609664 & $455\pm 16$ & -14.03 & 4.89 \\
 & B & $0.6333\pm 0.0068$ & $66.10\pm 0.57$ & $2.15\pm 0.11$ & 1679498552525334656\tablefootmark{$^{\ddagger}$} & - & - & - \\
 & \textit{Gaia} & 1.66 & 82.25 & 0.36 & 1679498552525334656 & $456\pm 25$ & -13.77 & 6.88 \\ \hline

TOI-1331 & A & - & - & - & 2171356454931124864 & $400.2\pm 1.9$ & 0.4 & -13.16 \\
 & B & $4.1999\pm 0.0016$ & $-68.090\pm 0.049$ & $2.380\pm 0.019$ & 2171356454931124992
 & $657.9 \pm 5.5$ & -1.64 & 2.97 \\ \hline

TOI-1450\tablefootmark{$^{\dagger\dagger}$} & A & - & - & - & 2155540255030594560 & $22.4425\pm 0.0079$ & 69.99 & 160.8 \\
 & B & $3.1709\pm 0.0032$ & $36.305\pm 0.025$ & $3.753\pm 0.036$ & 2155540255030594688\tablefootmark{$^{\ddagger}$} & - & - & - \\ 
 & \textit{Gaia}  & $4.77$ & $124.9$ & $3.8$ & 2155540255030594688 & $22.461\pm 0.013$ & 83.3 & 143.8 \\ \hline

TOI-1452\tablefootmark{$^{\dagger}$} & A & - & - & - & 2264839957167921024 & $30.504\pm 0.013$ & 7.8 & -74.08 \\
 & B & $3.0900\pm 0.0015$ & $-3.162\pm 0.016$ & $0.483\pm 0.016$ & 2264839952875245696 & $30.496\pm 0.013$ & 6.85 & -82.22 \\ \hline

TOI-1453\tablefootmark{$^{\dagger\dagger}$} & A & - & - & - & 1433062331332673792 & $78.887\pm 0.064$ & -70.97 & -0.09 \\
 & B & $1.8503\pm 0.0064$ & $56.04\pm 0.12$ & $4.354\pm 0.066$ & 1433062331333157632\tablefootmark{$^{\ddagger}$} & - & - & - \\ 
 & \textit{Gaia}  & $3.02$ & $110.7$ & $5.2$ & 1433062331333157632 & $78.53\pm0.42$ & 74.8 & 2.3 \\ \hline

TOI-1654\tablefootmark{$^{\dagger\dagger}$} & A & - & - & - & 2255021730651241728 & $292.6\pm 1.9$ & 6.02 & -19.17 \\
 & B & $1.0360\pm 0.0019$ & $81.33\pm 0.25$ & $0.938\pm 0.017$ & 2255021730648557184\tablefootmark{$^{\ddagger}$} & - & - & - \\ 
 & \textit{Gaia} & $2.6$ & $86.9$ & $0.81$ & 2255021730648557184 & $284.7 \pm 2.6$ & 4.8 & -18.9 \\ \hline

TOI-2072 & A & - & - & - & 1079565834014085632 & $39.018\pm 0.029$ & -184.97 & -87.24 \\
 & B & $2.9558\pm 0.0018$ & $-134.667\pm 0.018$ & $1.068\pm 0.020$ & - & - & - & - \\ 
 &  \textit{Gaia} & $9.3$ & $102$ & $0.92$ & 1079565834013188864 & $39.132\pm0.029$ & -184.6 & -92.0 \\ \hline

TOI-5128\tablefootmark{$^{\dagger}$} & A & - & - & - & 588253843440407808 & $192.9\pm 1.2$ & -17.34 & -14.67 \\
 & B & $1.4238\pm 0.0012$ & $3.808\pm 0.023$ & $2.324\pm 0.011$ & 588253843440048768 & $203.5\pm 1.9$ & -18.25 & -14.39 \\ \hline

TOI-5377 & A & - & - & - & 1040790426885870976 & $268\pm 40$ & -37.62 & -53.21 \\
 & B & $0.1975\pm 0.0013$ & $57.44\pm 0.28$ & $0.6845\pm 0.0099$ & - & - & - & - \\ \hline

TOI-5474 & A & - & - & - & 3450525873195051136 & $710.3\pm 7.1$ & -0.6 & -7.67 \\
 & B & $1.3339\pm 0.0036$ & $101.32\pm 0.55$ & $2.395\pm 0.046$ & 3450525877490978176\tablefootmark{$^{\ddagger}$} & - & - & - \\
 & \textit{Gaia} & 1.65 & 98.57 & 1.83 & 3450525877490978176 & - & - & - \\
\hline
\end{tabular}
\tablefoot{
\tablefoottext{$^{\dagger}$}{Companions detected by AstraLux are also confidently detected with \textit{Gaia} DR3, and their proper motions and distances point to them being physically bound to the main target.}
\tablefoottext{$^{\dagger\dagger}$}{The companions detected by AstraLux do not fully match the position of a nearby \textit{Gaia} DR3 source that is likely physically bound to the main target. This can either be because the AstraLux detection corresponds to another source or because we are detecting the binary orbital motion of the \textit{Gaia} source (see, e.g., the case of TOI-1450).}
\tablefoottext{$^{\ddagger}$}{\textit{Gaia} DR3 source ID to which the source could potentially be associated.}
}
\end{table*}

\subsection{Sensitivity curves}
\label{sec:sensitivity}
\begin{figure}[H]
\centering
\includegraphics[width=0.48\textwidth{}]{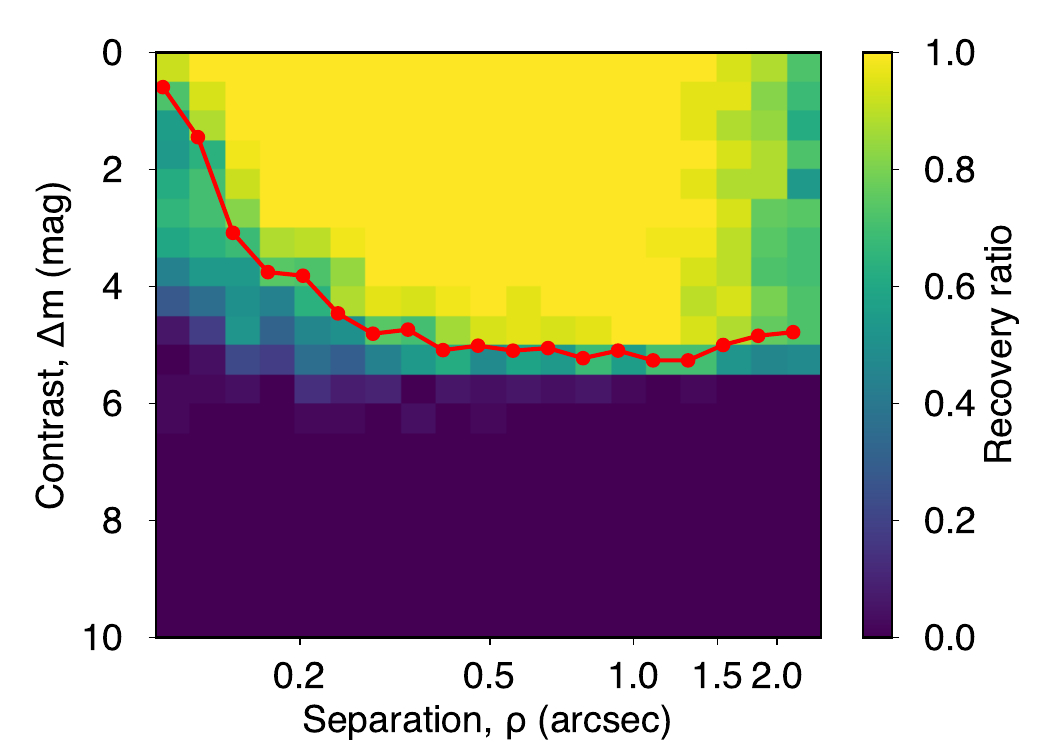}
\caption{Example of an injection-recovery matrix for one of our targets (TOI-1276). The color code indicates the rate of injected sources recovered. The red line represents the 70\% recovery contour assumed to be the sensitivity curve (see Sect.~\ref{sec:sensitivity}). }
\label{fig:matrix}
\end{figure}

Once the position of the target was identified, we could compute the sensitivity limits of our image. This was performed through a process similar to that described in \cite{lillo-box12}, Section 2.3.6. In this case, we used the PSF profile determined in the previous step and performed an injection-recovery analysis by artificially adding scaled profiles at different separations and position angles around the image. To that end, we used an array of 20 separations ($\rho$) between 0.1 arcsec and the minimum separation from the target to the edges of the image. For the contrast magnitude, we used an array of 20 values between $\Delta m=0$~mag and $\Delta m=10$~mag. For each pair of values \{$\rho$, $\Delta m$\}, we injected a source at 100 different position angles around the target. We then proceeded to the source identification of the injected signal as described in the previous section. We checked if the injected source was detected and stored this flag in a 20$\times$20-sized matrix, counting the number of recovered sources per separation-contrast bin. This is shown in Fig.~\ref{fig:matrix}. We set a 70\% recovery of the injected sources with a 5$\sigma$ significance as our criterion to define the sensitivity contour. Figure~\ref{fig:matrix} shows an example of the injection-recovery matrix and the correspondingly derived sensitivity limits (red line). The color in each separation-contrast bin in the matrix represents the recovery ratio (i.e., the percentage of injected sources detected by our method). We computed this sensitivity limit for all of our AstraLux images. 

\subsection{Likelihood of undetected blended sources}

\begin{figure}[H]
\centering
\includegraphics[width=0.48\textwidth{}]{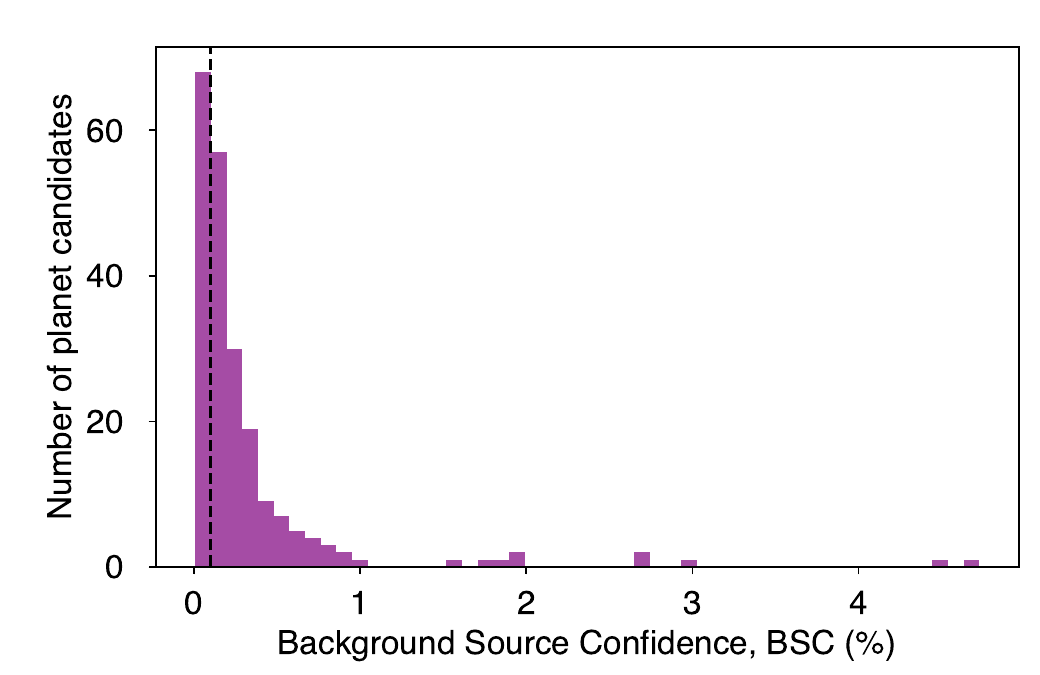}
\caption{Histogram of the BSC probability calculated after including the sensitivity of our high spatial resolution images for all TOIs observed in this work. {The dashed line marks where the BSC is equal to $0.1$\%}.}
\label{fig:BSCresult}
\end{figure}

For the sample of 164 isolated targets, we computed the probability of additional sources with capabilities of mimicking the given planetary transit (with the maximum magnitude contrast corresponding to that from Eq.~\ref{eq:contrast}) remaining undetected in our images. This is the metric that we call the blended source confidence (BSC; see \cite{lillo-box14b}). To that end, we used the \texttt{bsc} algorithm \citep{lillo-box14b}. This algorithm uses the TRILEGAL\footnote{\url{http://stev.oapd.inaf.it/cgi-bin/trilegal}} software \citep{trilegal} to obtain a population of Milky Way stars in the direction of the target to estimate the probability of undetected blended sources. By using this and the contrast curves of our high spatial resolution images presented in Sect.~\ref{sec:sensitivity}, we constrained the parameter space and provided the probability of the presence of blended undetected companions within 2 arcsec from the target. For specific details about the methodology, we refer the interested reader to \cite{lillo-box14b}. The probability of each target having undetected companions that are able to mimic the planetary transit is provided in the second-to-last column of Table~\ref{tab:BSCresults}. We note that this probability already provides a good proxy for the potential validity of the planet candidates. A histogram of the BSC values for all sources is shown in Fig.~\ref{fig:BSCresult}. The vast majority of the planet candidates (95\% of the sample) have BSC probabilities below 1\%, and around one-third of the candidates show a probability below $0.1$\% that other blended sources could mimic their transit signals.

\subsection{Centroid analysis}

In order to further check the validity of the planetary nature of candidates around isolated TOIs (as seen by our AstraLux images and within their sensitivity limits) that still have a planet candidate disposition, we performed a centroid analysis. {This yielded a probability of the transit actually happening on the target and not being contamination from nearby eclipsing binaries. A combination of a high probability of the transit occurring on-target, the absence of nearby companions through our high spatial resolution images, and a low probability of blended sources (i.e., low BSC) based on those images increases the chances of the planet candidates being actual planets and hence can help in prioritizing radial velocity efforts to infer their masses and confirm their planetary nature.} This test was also run for all the targets in the survey.

We used a modified version of the TESS positional probability software (\texttt{tpp}\footnote{\url{https://github.com/ahadjigeorghiou/TESSPositionalProbability}}) described in \cite{tpp} to compute the probability of each \gaia\-detected source in the field around the target as being the true source of the transiting signal. The method is based on comparing the observed centroid offsets to those that would be produced if the transit were to occur in other \textit{Gaia} sources surrounding the target. The resulting probability for each of the targets in our sample hosting the transiting signal is presented in the last column of Table~\ref{tab:BSCresults}. 

{First, to check the reliability of \texttt{tpp}, we used the sample of 26 already confirmed exoplanets}. Among them, 22 have $tpp$s above 80\%, indicating that they are the true source of the transit. However, in four cases, this probability is well below this threshold. The planet TOI-1611b (${tpp}=58$\%) was confirmed through radial velocities by \cite{heidari22}. Hence, since our AstraLux images do not show any companion within the fiber aperture of the instruments used (SOPHIE, HIRES, APF, typically 1 arcsec), the planet is safe from not being the primary source of the transit. The planet TOI-2260b (${tpp}=51$\%) was statistically validated by \cite{giacalone22}, and its nearby companions have been identified {with the aid of additional ground-based photometric data} as eclipsing binaries in the validation paper. Planet b in TOI-561 (${tpp}=62$\%) was confirmed by \cite{lacedelli21}, while its planet f (${tpp}=54$\%) was identified as a false positive by \cite{weiss21}. {Overall, this shows that \texttt{tpp} is able to assign large positional probabilities to confirmed planets}.

Among the seven candidates classified as eclipsing binaries (TOI disposition being EB; see Sect.~\ref{sec:Observations}), two (TOI-1157 and TOI-460) have a 100\% probability that the eclipses occur on the observed target. {Since no additional nearby companions have been found in our AstraLux images, we can confirm in these cases that the targets are indeed eclipsing binaries.} For each of the remaining five candidates, the probability of the eclipses occurring on the observed target is below 20\%, thus showing that the eclipsing binary is a nearby companion contaminating the light curve of the TOI. {None of these five candidates show companions in our AstraLux images, hence indicating that the source of the eclipses is a companion beyond 3 arcsec.}

When it comes to the 177 planet candidates still not verified (TOI disposition being PC), we found $tpp$ values above 95\% in 77 of them{, among which 74 show no additional sources in our AstraLux images. These are then optimal candidates for subsequent follow-up programs with radial velocity monitoring. In the remaining three candidates of the 77 (TOI-680, TOI-1297, and TOI-5128),} we detected close ($<2.2$~arcsec) companions, and only in one case (TOI-5128) is the companion in the \gaia\ catalog. Hence, since the \texttt{tpp} software relies on \gaia\-detected sources, for the other two, we cannot trust the $tpp$, as the algorithm does not take into account the presence of the new companions. {This clearly highlights the complementary relationship between statistical centroid analysis and high spatial resolution images.}


\begin{table*}
 \setlength{\extrarowheight}{3pt}
 \centering
\caption{Results from the analysis of the AstraLux images, including the contrast magnitudes at different separations, the BSC, and the $tpp$. A complete version of the table is included in the electronic version of this paper.}
\label{tab:BSCresults}
\begin{tabular}{ccccccccc}
 \hline 
 \hline 
TOI & planet & R$_{\rm p}$ & Disp. &$\Delta m(0.5^{\prime \prime})$ & $\Delta m(1^{\prime \prime})$ & $\Delta m(1.5^{\prime \prime})$ & BSC & $tpp$ \\
 & & (R$_{\oplus}$)& & mag. & mag. & mag. & \% & \% \\
 \hline 
TOI-1001 & .01 & 11.2 & PC & 4.07 & 3.87 & 3.8 & 0.177 & 32.8 \\
TOI-1131 & .01 & 12.1 & PC & 3.55 & 3.39 & 3.26 & 0.297 & 62.7 \\
TOI-1136 & .01 & 5.6 & PC & 4.4 & 4.35 & 4.28 & 0.027 & 100.0 \\
TOI-1136 & .02 & 2.5 & PC & 4.4 & 4.35 & 4.28 & 0.068 & 76.2 \\
TOI-1136 & .03 & 2.7 & PC & 4.4 & 4.35 & 4.28 & 0.058 & 100.0 \\
TOI-1136 & .04 & 2.5 & PC & 4.4 & 4.35 & 4.28 & 0.066 & 60.6 \\
TOI-1154 & .01 & 2.1 & PC & 4.22 & 4.16 & 3.77 & 0.189 & 89.3 \\
TOI-1157 & .01 & 35.8 & EB & 3.77 & 3.72 & 3.62 & 0.034 & 100.0 \\
TOI-1169 & .01 & 17.5 & PC & 3.9 & 3.87 & 3.79 & 0.015 & 38.6 \\
TOI-1174 & .01 & 2.4 & PC & 4.94 & 4.79 & 4.67 & 0.12 & 100.0 \\
TOI-1180 & .01 & 3.0 & PC & 4.69 & 4.76 & 4.78 & 0.114 & 97.9 \\
TOI-1184 & .01 & 2.3 & PC & 3.85 & 3.79 & 3.7 & 0.149 & 100.0 \\
TOI-1201 & .01 & 2.2 & PC & 3.98 & 3.79 & 3.68 & 0.111 & 73.8 \\
... & ...& ...& ...& ...& ... \\
 \hline 
\end{tabular}
\end{table*}

\section{Discussion and conclusions}
\label{sec:conclusions}

In this work, we have presented the release of high spatial resolution images of 185 TESS mission planet candidate host stars from a large survey using the AstraLux instrument at Calar Alto Observatory. We focused on stars hosting small planets ($R_p<4R_{\oplus}$), as these are the main focus of the TESS mission (the so-called LOSRs). Among other information, we provide sensitivity limits and derived the probability of undetected sources capable of mimicking the planetary signals being present for the 143 planet candidates, fulfilling this primary goal of the TESS space-based mission. Additionally, planets from other parts of the parameter space (including Neptunes in the hot-Neptune desert, planets around intermediate-mass stars, and planets around evolved stars) are also included in our sample. 

The results of the survey show {companions for 20 out of the 185 stars observed. Among them, 13 (7\%) show bright companions closer than {2.2 arcsec}}. These companions can certainly contaminate the light curves of the actual planet candidate hosts and thus bias the determination of the planet properties or even mimic planetary signals. Interestingly, among the 13 companion sources found, six of them are not detected by \gaia, and hence the TESS pipeline has not corrected their light curves from contamination due to these blended sources, thus biasing the properties of the planet candidates. This is a critical aspect for future missions such as PLATO and demonstrates the need for high spatial resolution images of planet candidates detected by space-based (large-aperture) photometers, even in the \gaia\ and post-\gaia\ era. 

Based on our high spatial resolution images, we have computed the probability of the presence of undetected sources with the capability of mimicking the planet transit in our AstraLux images. The results show that this probability is below 1\% for the vast majority of the observed sample, thus potentially eliminating other sources (apart from those of the \gaia\ DR3 catalog) as the origin of the transits. For all the targets in our sample, we also computed the $tpp$, which is the probability that the assigned target is the true source of the transit signal. The combination of a high $tpp$ and a low BSC implies that the transiting candidate is very likely to be of planetary nature. Hence, the information provided in this paper can be used as a way to prioritize subsequent follow-up (mostly radial velocitiy) observations toward the definitive confirmation of the planet candidates.

\begin{acknowledgements}
We thank the anonymous referee for their though revision of this manuscript that has improved its final quality.
J.L.-B. is partly funded by grants LCF/BQ/PI20/11760023 and Ram\'on y Cajal fellowship with code RYC2021-031640-I. 
A.C.-G., D.B., J.L.-B., M.M.-C., N.H. and O.B.-R. are also partly funded by the Spanish MICIU/AEI/10.13039/501100011033 grant PID2019-107061GB-C61. J.L.-B. is also funded by the MICIU/AEI/10.13039/501100011033 and NextGenerationEU/PRTR grant CNS2023-144309. 
I.M.'s research is funded by grants PID2022-138366NA-I00, by the Spanish Ministry of Science and Innovation/State Agency of Research MCIN/AEI/10.13039/501100011033 and by the European Union, and by a Ram\'on y Cajal fellowship RyC2019-026992-I. This work has made use of the following python modules: 
This research made use of \texttt{astropy}, (a community-developed core Python package for Astronomy, \citealt{astropy:2013,astropy:2018}), \texttt{SciPy} \citep{scipy}, \texttt{matplotlib} (a Python library for publication quality graphics \citealt{matplotlib}), \texttt{astroML} \citep{astroML}, and \texttt{numpy} \citep{numpy}.
This research has made use of NASA's Astrophysics Data System (ADS) Bibliographic Services. 
This research has made use of the SIMBAD database, operated at CDS. 
This research has made use of the NASA Exoplanet Archive, which is operated by the California Institute of Technology, under contract with the National Aeronautics and Space Administration under the Exoplanet Exploration Program.
This research has made use of the Exoplanet Follow-up Observation Program (ExoFOP; DOI: 10.26134/ExoFOP5) website, which is operated by the California Institute of Technology, under contract with the National Aeronautics and Space Administration under the Exoplanet Exploration Program.
\end{acknowledgements}

%
%

\bibliographystyle{aa} 
\bibliography{biblio2} 

\end{document}